\documentclass[sigconf]{acmart}

\copyrightyear{2020}
\acmYear{2020}
\setcopyright{acmcopyright}\acmConference[RecSys '20]{Fourteenth ACM Conference on Recommender Systems}{September 22--26, 2020}{Virtual Event, Brazil}
\acmBooktitle{Fourteenth ACM Conference on Recommender Systems (RecSys '20), September 22--26, 2020, Virtual Event, Brazil}
\acmPrice{15.00}
\acmDOI{10.1145/3383313.3412237}
\acmISBN{978-1-4503-7583-2/20/09}

\usepackage{graphicx}
\usepackage{amsmath}

\usepackage{multirow}
\usepackage{float} 
\usepackage{subfigure} 
\usepackage{enumitem}   
\usepackage{amssymb}

\begin{document}
\title[Knowledge-Aware Document Representation for News Recommendations]{KRED: Knowledge-Aware Document Representation for News Recommendations}

\author{Danyang Liu} 
\affiliation{%
  \institution{University of Science and Technology of China} 
}
\email{ldy591@mail.ustc.edu.cn}  

\author{Jianxun Lian}
\authornote{Corresponding author.}
\affiliation{%
  \institution{Microsoft Research Asia} 
  }
\email{Jianxun.Lian@microsoft.com}

\author{Shiyin Wang}
\affiliation{%
  \institution{Tsinghua University} 
  }
\email{wangshiy16@mails.tsinghua.edu.cn}

\author{Ying Qiao}
\email{yiqia@microsoft.com}
\author{Jiun-Hung Chen}
\email{jiuche@microsoft.com}
\affiliation{%
  \institution{Microsoft Corp.} 
 }

\author{Guangzhong Sun}
\affiliation{%
  \institution{University of Science and Technology of China} 
}
\email{gzsun@ustc.edu.cn}   

\author{Xing Xie}
\affiliation{%
  \institution{Microsoft Research Asia} 
  }
\email{Xing.Xie@microsoft.com}

\newcommand{\ie}{\textit{i}.\textit{e}.}
\newcommand{\eg}{\textit{e}.\textit{g}.}

\renewcommand{\shortauthors}{Liu and Lian, et al.}

\begin{abstract}
News articles usually contain knowledge entities such as celebrities or organizations. Important entities in articles carry key messages and help to understand the content in a more direct way.
An industrial news recommender system contains various key applications, such as personalized recommendation, item-to-item recommendation, news category classification, news popularity prediction and local news detection. We find that incorporating knowledge entities for better document understanding benefits these applications consistently.
However, existing document understanding models either represent news articles without considering knowledge entities (e.g., BERT) or rely on a specific type of text encoding model (e.g., DKN) so that the generalization ability and efficiency is compromised.  
In this paper, we propose KRED, which is a fast and effective model to enhance arbitrary document representation with a knowledge graph. KRED first enriches entities' embeddings by attentively aggregating information from their neighborhood in the knowledge graph. Then a context embedding layer is applied to annotate the dynamic context of different entities such as frequency, category and position. Finally, an information distillation layer aggregates the entity embeddings under the guidance of the original document representation and transforms the document vector into a new one.
We advocate to optimize the model with a multi-task framework, so that different news recommendation applications can be united and useful information can be shared across different tasks.
Experiments on a real-world Microsoft News dataset demonstrate that KRED greatly benefits a variety of news recommendation applications. 
\end{abstract}



\maketitle
 
\section{Introduction}
\begin{figure}[t]
    \centering
    \includegraphics[width=0.5\textwidth]{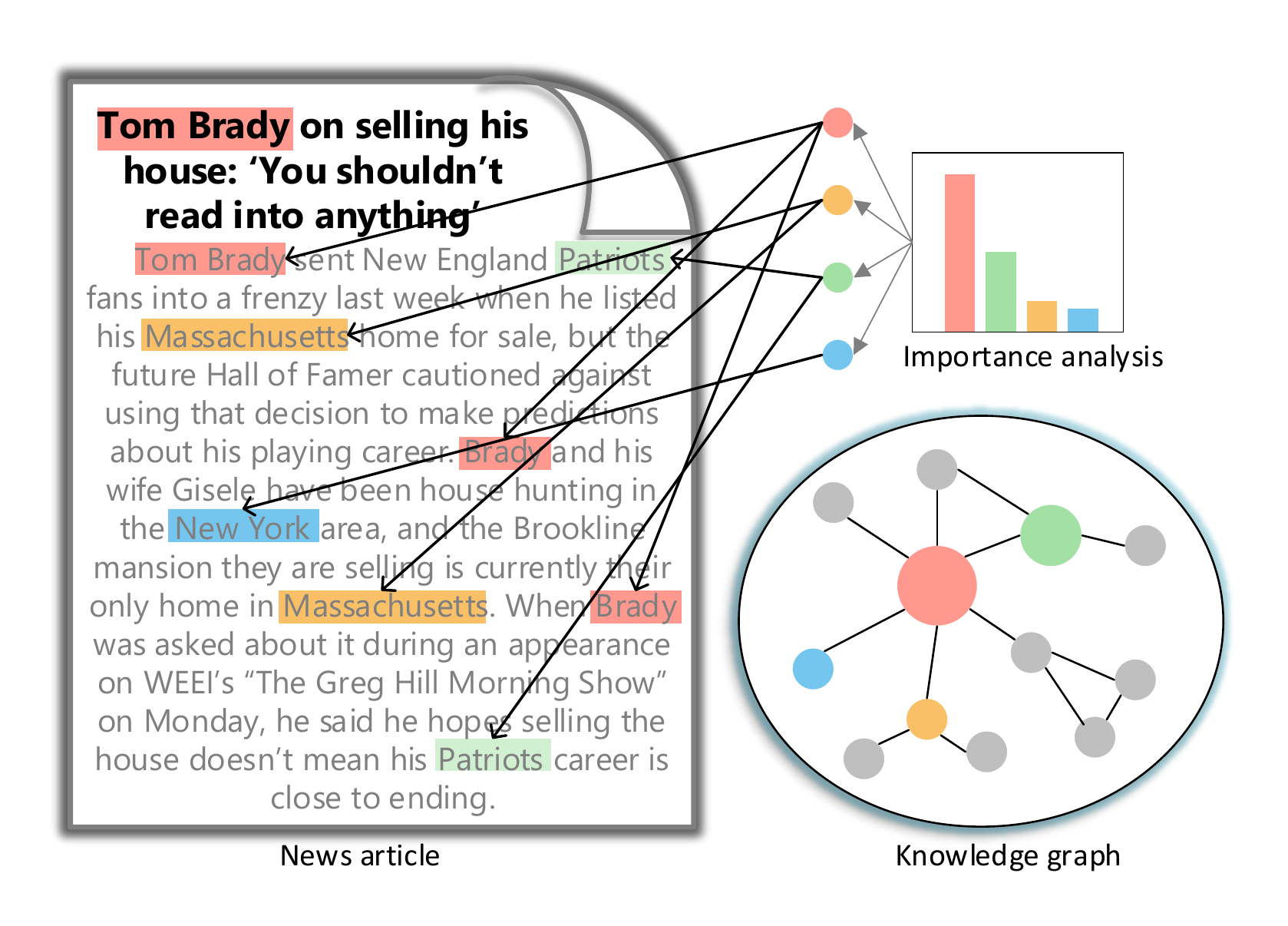}%
    \caption{An illustration of a real news article. An article can contain many entities, including those important and unimportant. The grasp of entities helps us understand the news content in a quicker and easier manner.}
    \label{fig:case_study}
\end{figure}

With the vigorous development of the Internet, online news has become an increasingly important source of daily news consumption for users. 
Considering that hundreds of thousands of news articles can be produced every day, recommender systems become critical for news service providers to improve user experience.  Distinguished from other recommendation domains such as movies, music and restaurants, news recommendation has three characteristics. First, news articles are highly time-sensitive. As revealed in \cite{Wang:2018:DDK:3178876.3186175}, about 90\% of news expire within just two days. Therefore, classical ID-based collaborative filtering methods are less effective in this situation. A deep understanding of news content is necessary. Second, news articles possess the accuracy, brevity and clarity characteristics. Thus, natural language understanding (NLU) models can be more effectively applied to produce high-quality document representations for user interest capturing, compared with other complex textual messages such as an unstructured webpage or abstruse poetry. Third, news articles may contain a few entities, e.g., celebrities, cities, companies or products. These entities are usually the key message conveyed by the article. 
Figure \ref{fig:case_study} shows a piece of real news, where multiple entities are mentioned in the article (for the sake of brevity, some of the entities are not highlighted). By considering the position, frequency, and relationships of the entities, the article can be better comprehended from the knowledge entities' perspective. Given that a user has clicked on this article, it makes more sense to recommend more news related to \textsl{Tom Brady} rather than \textsl{Massachusetts}. Since a knowledge graph contains comprehensive external information of entities, leveraging knowledge graph for better news document understanding became a promising research direction. 

Previous works in news recommendation manually extract features from news items \cite{Lian:2018:TBR:3304222.3304298} or extract latent representations via neural models \cite{Okura:2017:ENR:3097983.3098108,DBLP:conf/ijcai/WuWAHHX19}. These approaches ignore the importance of entities in an article. In the direction of integrating knowledge graphs for news recommendation, the most related work is DKN \cite{Wang:2018:DDK:3178876.3186175}. The key component of DKN is the knowledge-aware convolutional neural network (KCNN). However, DKN only takes news titles as input. While extending to incorporate the news body is possible, it will lead to inefficiency problems.  Meanwhile, DKN uses a Kim-CNN \cite{DBLP:conf/emnlp/Kim14} framework to fuse words and associated entities, which is not flexible and many more sophisticated models such as Transformers \cite{Vaswani:2017:AYN:3295222.3295349} cannot be applied. With the great success of BERT \cite{DBLP:conf/naacl/DevlinCLT19}, pretraining + finetuning has gradually become a new paradigm in the NLU domain. The pretraining step is conducted on large-scale datasets and is expensive. But once finished, the pretrained model can be shared so that people can enhance their models by finetuning the pretrained model with their specific tasks. Thus, we are curious to explore whether there is a framework which can integrate the information from knowledge graphs, while do not have restriction on the kind of original document representation?

To this end, we propose \textbf{KRED}, which stands for a \underline{K}nowledge-aware \underline{R}epresentation \underline{E}nhancement model for news \underline{D}ocuments. Given a document vector which can be produced by any type of NLU model (such as BERT \cite{DBLP:conf/naacl/DevlinCLT19}), KRED fuses knowledge entities included in the article and produces a new representation vector in a fast, elastic and accurate manner. The new document vector can be consumed by many downstream tasks. 
There are three layers in our proposed model, \ie an entity representation layer, a context embedding layer, and an information distillation layer. The original entity embedding is trained based on a knowledge graph using TransE \cite{Bordes:2013:TEM:2999792.2999923}. Motivated by the recent progress of the Knowledge Graph Attention network (KGAT) \cite{Wang:2019:KKG:3292500.3330989}, we refine an entity's representation with its surrounding entities. Considering that the context of entities can be complex and dynamic, we design a context embedding layer to incorporate different contextual information such as position, category and frequency. Lastly, since the importance of different entities depends on entities themselves, the co-occurrence of entities, and topics in the current article, we use an attentive mechanism to merge all entities into one fixed-length embedding vector, which can be concatenated with the original document vector and then be transformed into a new document vector. 

In this paper, we refer to \textbf{news recommendations} as a broad definition of a series of applications, which not only includes \textbf{personalized recommendation}, but also includes \textbf{item-to-item recommendation}, \textbf{news popularity prediction}, \textbf{new category prediction} and \textbf{local news detection}. An industrial recommender system usually provides various recommendation services, such as recommendations by personal interest, or recommendations by item popularity / location / category. These tasks share some common data patterns, e.g., users with similar interests trend to read news articles with similar topics (personalized recommendations), meanwhile similar users trend to read news with the same category (category classification). Bridging models of these tasks can enrich the data and help to regularize the models. We propose to jointly train these applications in a multi-task framework. Through extensive experiments on a real-world news reading dataset, we observe a significant performance gain, especially in low resources applications (such as local news detection, which has a low ratio of positive labels).
To summarize, we make the following contributions:
\begin{itemize}
    \item We propose a knowledge-enhanced document representation model, named KRED, to improve the news article representation in a fast, elastic, accurate manner. KRED is fully flexible and can enhance an arbitrary base document vector with knowledge entities. In the experiment section we demonstrate this merit with two different base document vector settings, i.e., LDA + DSSM and BERT.
    \item To fully utilize knowledge information, we propose three key components in KRED, including an entity representation layer, a context embedding layer, and an information distillation layer. Through ablation studies we verify that indeed each component contributes to the model.
    \item We highlight the necessity of connecting multiple news recommendation applications and propose to train the KRED in a multi-task framework. To the best of our knowledge, we are the first to propose a knowledge-aware model to serve various news recommendation applications. By conducting comprehensive experiments, we demonstrate that this joint training framework significantly improves the performance on various news recommendation tasks.
    \item We provide some visual case studies to illustrate in a user-friendly manner how KRED help to understand documents and generate document embeddings. 
\end{itemize}

\section{The Proposed Method}
\begin{figure*}[ht]
    \centering
    \includegraphics[width=1.0\textwidth,trim={0 0cm 0 0},clip]{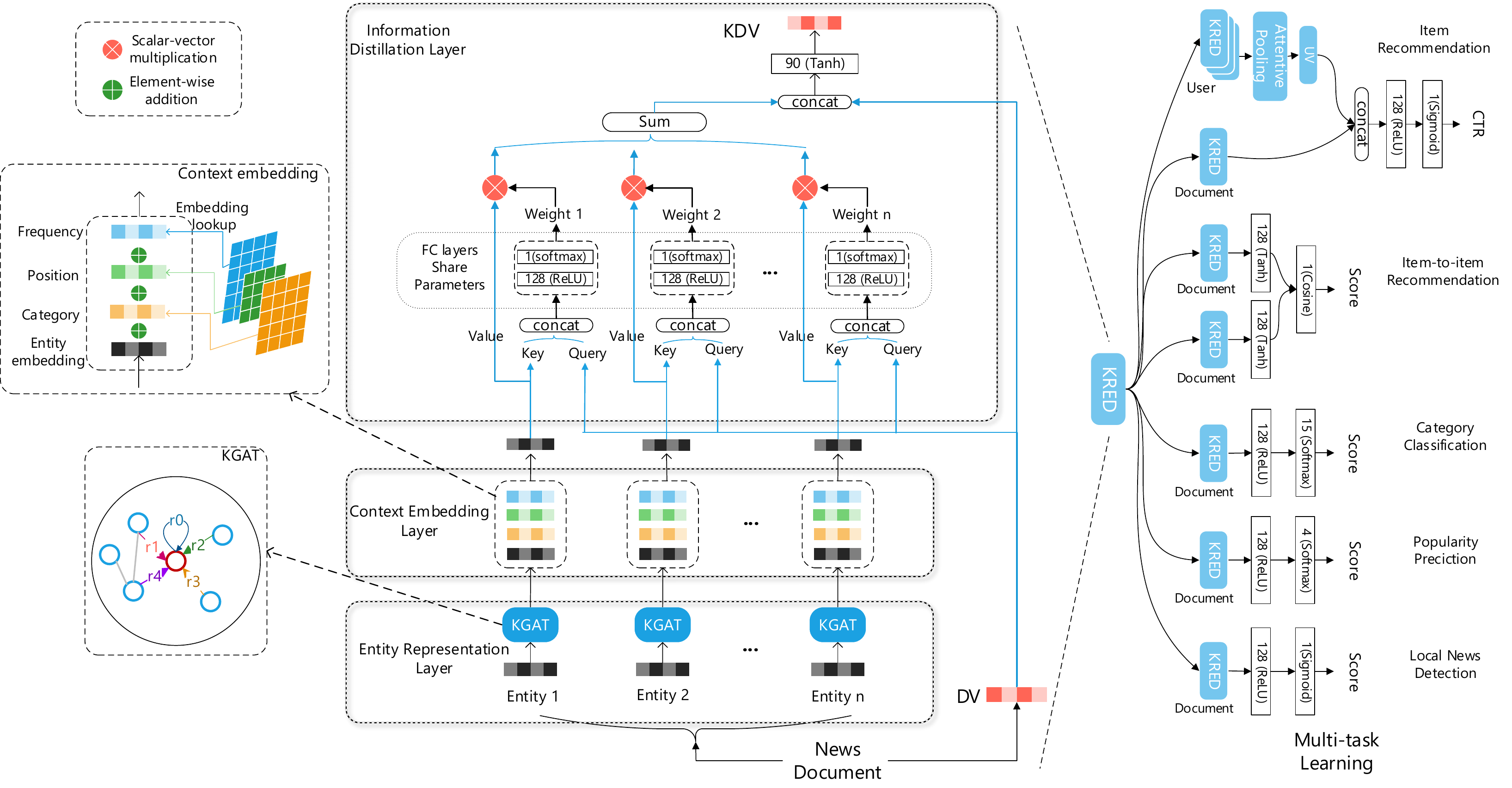}%
    \caption{An overview of the proposed \emph{KRED} model. DV indicates the (original) document vector. KDV indicates the knowledge-enhanced document vector produced by KRED. UV indicates the user vector.}
    \label{fig:overall_KDV}
\end{figure*}

We propose a \underline{K}nowledge-aware \underline{R}epresentation \underline{E}nhancement model for news \underline{D}ocuments (KRED), which is illustrated in Figure \ref{fig:overall_KDV}. Given an arbitrary Document Vector (DV, such as produced by BERT), denoted as $\mathbf{v}_d$, our goal is to produce a Knowledge-enhanced Document Vector (KDV), which can be consumed by various downstream applications. The benefits of our framework are three-fold. First, unlike DKN which relies on Kim-CNN, it has no constraint on which specific document representation method should be used. In the age of pretrained + finetuning paradigm, all kinds of pretrained models can be used for our framework. Second, it can utilize all the data included in the news document, e.g., title, body and meta-data. Third, it is fast because on one hand it fuses knowledge entities into a base document vector without re-computing the whole text strings; on the other hand, we do our best to select lightweight operations for the intermediate components. Next we introduce three key layers in KRED: an entity representation layer, a context embedding layer, an information distillation layer.

\subsection{Entity Representation Layer}\label{sec:entityrep}
 We assume that entities in news articles can be linked to corresponding entities in a knowledge graph. A knowledge graph is formulated as a collection of entity-relation-entity triples: $\mathcal{G}=\{(h,r,t)|h,t\in \mathcal{E}, r\in \mathcal{R}\}$, where $\mathcal{E}$ and $\mathcal{R}$ represent the set of entities and relations respectively, $(h,r,t)$ represents that there is a relation $r$ from $h$ to $t$. 
 We use TransE \cite{Bordes:2013:TEM:2999792.2999923} to learn embedding vectors $\mathbf{e} \in \mathbb{R}^d$ for each entity and relationship. TransE is trained on the knowledge graph and embeddings of entity/relationship are kept fixed as features for our framework.
 
Considering that an entity cannot only be represented by its own embeddings, but also partially represented by its neighbors, we exploit the idea of Knowledge Graph Attention (KGAT) Network \cite{Wang:2019:KKG:3292500.3330989} to produce an entity representation. Let $\mathcal{N}_h$ denote the set of triplets where $h$ is the head entity. An entity is represented by:
\begin{equation}
    \textbf{e}_{\mathcal{N}_h} = ReLU \Big( \mathbf{W_0} \big(\mathbf{e}_h \oplus \sum_{(h,r,t)\in\mathcal{N}_h} \pi(h,r,t)\textbf{e}_t \big) \Big)
\end{equation}
where $\oplus$ denotes the vector concatenation, $\mathbf{e}_h$ and $\textbf{e}_t$ are the entity vectors learned from TransE.  $\pi(h,r,t)$ is the attention weight that controls how much information the neighbor node need to propagate to the current entity, and it is calculated via a two-layer fully connected neural network:
\begin{equation}
    \pi_0(h,r,t) = \textbf{w}_2 ReLU\big(\textbf{W}_1(\textbf{e}_h \oplus \textbf{e}_r \oplus \textbf{e}_t) + \textbf{b}_1\big) + b_2 
\end{equation}
\begin{equation}
    \pi(h,r,t) = \frac{exp\big( \pi_0(h,r,t) \big)}{\sum_{(h,r',t')\in \mathcal{N}_h} exp\big(\pi_0(h,r',t')\big)}
    \label{eq:entity_soft}
\end{equation}
Here we use a softmax function to normalize the coefficients. We can take the idea of graph neural networks \cite{Ying:2018:GCN:3219819.3219890} to train the model with higher-order information propagation, where neighborhood aggregation is stacked for multiple iterations. However, it comes with a heavy computational cost and the performance gain is insignificant considering the accompanying model complexity. Therefore we only enable one-hop graph neighborhood, the TransE embedding vector $\textbf{e}$ is taken as node features and the trainable parameters are $\{\mathbf{W_0}, \textbf{W}_1, \textbf{w}_2, \textbf{b}_1, b_2\}$.
\subsection{Context Embedding Layer}\label{sec:contextembedding}
To reduce the computational cost, we avoid taking the whole original document as the model's input. An efficient way is to extract entities' conclusive information from the document. We observe that an entity may appear in different documents in various ways, such as position and frequency. The dynamic context heavily influences the importance and relevance of the entity in the article. We design three context embedding features to encode the dynamic context:

\noindent\textbf{Position Encoding}. 
The position means whether the entity appears in the title or body. In many cases, entities in the news title are more important than those that only appear in the news body. We add a position bias vector $\textbf{C}_{p_i}^{(1)}$ to the entity embeddings, where $p_i \in \{1,2\}$ indicates the position type of entity $i$. 

\noindent\textbf{Frequency Encoding}.
The frequency can tell the importance of entities to some extent. Thus, we create an encoding matrix $\textbf{C}^{(2)}$ for entity frequency. We count the occurrence frequency $f_i$ of each entity, use it as a discrete index to look up a frequency encoding vector $\textbf{C}^{(2)}_{f_i}$, then add it to the entity embedding vector. The upper bound of $f_i$ is set to 20.

\noindent\textbf{Category Encoding}
Entities belong to different categories, e.g., \textsl{Donald Trump} is a person, \textsl{Microsoft} is a company, \textsl{RecSys} is a conference. Explicitly revealing the category of entities helps the model understand content more easily and more accurately. Thus we maintain a category encoding matrix $\textbf{C}^{(3)}$. For each entity $i$ with type $t_i$, we add a category encoding vector $\textbf{C}^{(3)}_{t_i}$ to its embedding vector.

After the context embedding layer, for each entity $h$, its embedding vector as input for the next layer is a compound vector:
\begin{equation}
    \textbf{e}_{\mathcal{I}_h} =  \textbf{e}_{\mathcal{N}_h} + \textbf{C}_{p_h}^{(1)} + \textbf{C}^{(2)}_{f_h} + \textbf{C}^{(3)}_{t_h}
\end{equation}
where $+$ indicates the element-wise addition of vectors.
\subsection{Information Distillation Layer}
The final importance of an entity is not only determined by its own message, but also influenced by the other entities co-occurring in the article and the topics of the article. For instance, suppose there are two news articles related to a city A. The first article reports that a famous music star will play concerts in city A, while the second news reports a strong earthquake happened in city A. Obviously, the key entity in the former article is the celebrity, while in the latter it is the location. We use an attentive mechanism to merge all entities' information into one output vector. We follow the terminologies (query, key and value) in Transformer \cite{Vaswani:2017:AYN:3295222.3295349}. The original document vector $\mathbf{v}_d$  serves as the query. Both key and value are entity representation $\textbf{e}_{\mathcal{I}_h}$. The attention weight is computed as\footnote{For simplicity of notation, we reuse some notations in Section \ref{sec:entityrep}. Please note that they are different parameters.}:
\begin{equation}
    \pi_0(h,v) = \textbf{w}_2 ReLU\big(\textbf{W}_1(\textbf{e}_{\mathcal{I}_h} \oplus \mathbf{v}_d) + \textbf{b}_1\big) + b_2 
\end{equation}
\begin{equation}
    \pi(h,v) = \frac{exp\big( \pi_0(h,v) \big)}{\sum_{t\in \mathcal{E}_v} exp\big(\pi_0(t,v)\big)}
    \label{eq:entity_soft02}
\end{equation}
\begin{equation}
    \textbf{e}_{\mathcal{O}_h} = \sum_{h\in\mathcal{E}_v} \pi(h,v)\textbf{e}_{\mathcal{I}_h}
\end{equation}
Where $\mathcal{E}_v$ represents the entity set of document $v$. Then the entity vector and the original document vector is concatenated and goes through one fully-connected feed-forward network:
\begin{equation}
    \mathbf{v}_k = Tanh\big( \textbf{W}_3 (\textbf{e}_{\mathcal{O}_h} \oplus \mathbf{v}_d) + \textbf{b}_3\big)
\end{equation}
$\mathbf{v}_k$ is the Knowledge-aware Document Vector (KDV). It is interesting to point out that we didn't use the self-attention encoder or multi-head attention mechanism, because via experiments we observe that these components didn't lead to a significant improvement. A possible reason is that relationships between entities in one news document are not as complex as raw text in NLU, so adding complex layers is useless while bringing unnecessary computation cost.

\subsection{Multi-Task Learning}
\label{multi-task}
In industrial recommendation systems, there are a few other important tasks in addition to personalized item recommendation (aka, user2item recommendation), such as item-to-item recommendation (item2item), news popularity prediction, news category classification, and local news detection \footnote{Local news represent news articles which report events that happened in a local context that would not be an interest of another locality}. We find that these tasks share some knowledge patterns and their data can complement each other. Thus we design a multi-task learning approach to train the KRED model, as illustrated in Figure \ref{fig:overall_KDV}. 
It's intuitive to understand why a multi-task learning approach is better than to train models separately using task-specific data. Take the news category classification task for example. Our labelled dataset for category classification is limited to the size of our document corpus. However, the user-item interaction dataset is much larger. Considering that similar users trend to read news with similar topics, transferring the collaborative signals from user-item interactions can greatly improve the learning of news category classification.

In the multi-task framework, all the tasks share the same KRED as the backbone model, but on top of KRED they include a few task-specific parameters as the predictors. Among these tasks, only the item recommendation task involves a user component. All the other tasks only depend on the document component. Thus, in this section we introduce two kinds of predictors and loss functions.  

\noindent\textbf{Predictors}. For user2item recommendations, the input vectors include a user vector $\mathbf{u}_i$ and a document vector $\mathbf{v}_j$ (for brevity we remove the knowledge-aware subscript $k$), the predictive score is :
\begin{equation}
    \hat{y}(i,j)= g( \textbf{u}_i \oplus \textbf{v}_j )
\end{equation}
where $g$ denotes the predictive function which includes a one-layer neural network. We adopt a widely-used attentive merging approach to generate the user vector $\mathbf{u}_i$. Basically, we feed the document vector of his/her historical clicked articles into a one-layer neural network to get an attention score. Then we merge document vectors weighted by the attention scores to get the user vector $\mathbf{u}_i$. For the item2item recommendation task, we add one neural layer (with 128 dimension and tanh activation) to transform the document vector, then the relationship of two documents is measured by the cosine similarity. For all other tasks, the input vector is only a document vector $\mathbf{v}_j$, so the predictive score is $ \hat{y}(j)= g(\mathbf{v}_j)$. The predictors are illustrated in Figure \ref{fig:overall_KDV}.

\noindent\textbf{Loss functions}. For  user2item (item2item) recommendations, we exploit a ranking-based loss function for optimization. Given a positive user-item (item-item) pair $(i,j)$ in the training set, we randomly sample 5 items to compose negative user-item (item-item) pairs $(i,j')$. The training process maximizes the probability of positive instances:
\begin{equation}
    P(j|i)=\frac{exp(\gamma \hat{y}(i,j))}{\sum_{j'\in \mathcal{J}} exp(\gamma \hat{y}(i,j'))}
\end{equation}
where $\mathcal{J}$ denotes the candidate set which contains one positive item and five negative items, $\gamma$ is a smoothing factor for softmax function, which is set to 10 according to a held-out validation dataset. Thus the loss function we need to minimize becomes:
\begin{equation}
    \mathcal{L}_{rec}=-log \prod_{(i,j) \in H} P(j|i) + \lambda ||\Theta||
\end{equation}
where $H$ denotes the set of user-item click history, $\Theta$ denotes the set of trainable parameters, and $\lambda$ is the regularization coefficient. 

For all other tasks, since they belong to multi-class classification or binary classification problems, we take cross entropy as the objective function:
\begin{align}
    \mathcal{L}_{cls}= \sum_{c=1}^{M}  - y_{j,c} log(\hat{y}(j,c)) + \lambda ||\Theta||
\end{align}
where $y_{j,c}=1$ when the label of $y(i)$ is $c$, otherwise $y_{j,c}=0$. $M$ denotes the maximum number of labels. Note that binary classification problem can be regarded as a special case of a multi-class classification problem. To avoid involving new hyper-parameters to combine different tasks' loss function, we use a two-stage approach to conduct the multi-task training. In the first stage, we alternately train different tasks every few mini-batches. In the second stage, we only include the target task's data to finalize a task-specific model.

\section{Experiments}
\subsection{Dataset and Settings}
We use a real-world industrial dataset provided by Microsoft News \footnote{\url{https://news.microsoft.com}} (previously known as MSN News\footnote{\url{https://www.msn.com/en-us/news}}). We collect a set of impression logs ranging from Jan 15, 2019 to Jan 28, 2019. For personalized recommendation task, the first week is used for the training and validation set, while the latter week is used for the test set. In order to build user profiles, we collect another two weeks of impression logs prior to the training date and aggregate each user's behaviors for user modeling. We filter out users who clicked on less than 5 articles in the profile building period. After filtering, in the instance set (\ie  training + valid + test set) there are in total 665,034 users, 24,542 news articles, 1,590,092 interactions. Note that these impressions are collected from the homepage of our news site, on which all of the articles are reviewed manually by professional editors and are of high quality. Therefore, the number of articles is not very large. The average number of words in one document is 701. We use an internal tool to link news entities to an industrial knowledge graph Satori \cite{gao2018building}. We only keep the entities whose linking confidence scores are higher than 0.9. We process the original knowledge graph with the technique of News Graph \cite{DBLP:conf/cikm/LiuBLZSWX19}. On average, one document contains 24 entities and their 1-hop neighborhood in the knowledge graph cover 3,314,628 entities, 1,006 relations and 71,727,874 triples. An open-source version of the Microsoft News dataset is MIND \cite{wu-etal-2020-mind}.

In this paper, we consider five important tasks in news recommender systems, including personalized recommendations, item-to-item recommendations, new popularity prediction, news category classification and local news detection. For the news popularity prediction task, we split the news articles into four equal-sized groups according to their clicked volume, so each group indicates a certain level of popularity (known as mediocre, popular, super popular and viral). Local news detection is an imbalanced binary classification task, whose positive ratio is only 12\% in our dataset.  In the news category classification task, we have 15 top-level categories in the news corpus.

\noindent\textbf{Baselines}. We aim at learning knowledge-aware document embeddings and compare KRED with several groups of methods:
\begin{itemize}[leftmargin=*]
\item The first group includes different types of ordinary document representation models which do not use knowledge graphs. We want to highlight how our model can improve these models by injecting knowledge information. In this direction, we choose LDA+DSSM \cite{Blei:2003:LDA:944919.944937,Huang:2013:LDS:2541176.2505665}
and BERT (BERT$_{base}$, finetuned to our dataset) as the base document vector, denoted as \textbf{DV$_{LDA+DSSM}$} and \textbf{DV$_{BERT}$}. We also provide a simple way to inject knowledge information as a baseline, which is an attentive pooling of entity embeddings and then concatenate with the document vector (denoted as \textbf{DV$_{LDA+DSSM}$ + entity} and \textbf{DV$_{BERT}$ + entity}).
In addition, \textbf{NAML} \cite{DBLP:conf/ijcai/WuWAHHX19} is a strong baseline which uses an attentive multi-view learning model to learn news representations from multiple types of news content, such as titles, bodies and categories (but it didn't consider knowledge entities).
\item The second group includes several state-of-the-art models which explicitly use a knowledge graph in document modeling. We compare with them to demonstrate the effectiveness of the knowledge fusion mechanism of our proposed model. In this direction, we have \textbf{DKN} \cite{Wang:2018:DDK:3178876.3186175}, \textbf{STCKA} \cite{DBLP:conf/aaai/ChenHLXJ19} and \textbf{ERNIE} \cite{DBLP:conf/acl/ZhangHLJSL19} as the baseline models. For our model KRED, we conduct experiments on both base configurations of LDA+DSSM and BERT, which are denoted as \textbf{KRED$_{LDA+DSSM}$} and \textbf{KRED$_{BERT}$}, respectively. \textbf{KRED$_{BERT}$  single-task} indicates a variant in which we remove the multi-task training mechanism and train toward the target task directly. Note that both BERT and ERNIE are finetuned on our news dataset.
\item On personalized recommendations task, we compare with a third group of baselines: \textbf{FM} \cite{rendle2010factorization} and \textbf{Wide \& Deep} \cite{DBLP:journals/corr/ChengKHSCAACCIA16}. They are two strong baselines in the field of recommender systems. Different from the aforementioned methods which automatically learn high-level document representations, they take hand-craft features as inputs and learn feature interactions for making predictions. For these two baselines, entities are fed as additional features into the model. 
\end{itemize}

\begin{table*}[t]
\setlength{\abovecaptionskip}{0.2cm} 
\caption{Performance on the personalized user2item task. Numbers in \textbf{black bold} indicate a significantly ($p$-value\textless 0.01) improvement over the best baseline method (Baseline methods include all models except KRED$_*$ series).}%
\label{tab:overall_auc}
\centering
\begin{tabular}{c|c|c|c|c|c|c|c|c|c}
 \specialrule{1.5pt}{1pt}{1pt}
\multirow{2}{*}{} &  \multicolumn{8}{c|}{AUC}  & NDCG@10  \\  
\cline{2-10} 
                          &    Day1               & Day2               & Day3             & Day4             & Day5             & Day6             & Day7             & Overall  & Overall        
                          \\ 
\hline
FM                         & 0.6561           & 0.6603           & 0.6517           & 0.6598           & 0.6582           & 0.6402           & 0.6612           & 0.6577   & 0.2469         
\\ 
\hline
Wide \& Deep                   & 0.6628           & 0.6701           & 0.6579           & 0.6646           & 0.6635           & 0.6487           & 0.6733           & 0.6639   & 0.2503         
\\ 
\hline
NAML                       & 0.6741           & 0.6813           & 0.6735           & 0.6795           & 0.6741           & 0.6703           & 0.6803           & 0.6760  & 0.2579          
\\ 
\hline\hline
DKN                        & 0.6657           & 0.6718           & 0.6688           & 0.6727           & 0.6625           & 0.6599           & 0.6663           & 0.6686    & 0.2515        
\\ 
\hline
STCKA                      & 0.6723           & 0.6816           & 0.6674           & 0.6812           & 0.6728           & 0.6622           & 0.6721           & 0.6735    & 0.2588        
\\ 
\hline
ERNIE                      & 0.6780           & 0.6841           & 0.6776           & 0.6871           & 0.6732           & 0.6741           & 0.6841           & 0.6813   & 0.2607         
\\ 
\hline\hline
DV$_{LDA+DSSM}$            & 0.6591           & 0.6681           & 0.6552           & 0.6628           & 0.6646           & 0.6408           & 0.6711           & 0.6628    & 0.2485        
\\ 
\hline
DV$_{LDA+DSSM}$ + entity     & 0.6639           & 0.6769           & 0.6703           & 0.6787           & 0.6648           & 0.6521           & 0.6715           & 0.6720    & 0.2527        
\\ 
\hline
KRED$_{LDA+DSSM}$  single-task        &0.6808 &0.6806 &0.6890 &0.6882 &0.6841 &0.6684 &0.6779 &0.6831  & 0.2608
           \\ 
\hline
KRED$_{LDA+DSSM}$     &0.6873 &0.6894 &0.6905 &0.6897 &0.6858 &0.6734 &0.6798 &0.6871  & 0.2671
               \\ 
\hline\hline
DV$_{BERT}$                & 0.6758           & 0.6802           & 0.6763           & 0.6828           & 0.6721           & 0.6728           & 0.6807           & 0.6792    & 0.2590        
\\ 
\hline
DV$_{BERT}$ + entity         & 0.6813           & 0.6904           & 0.6809           & 0.6962           & 0.6761           & 0.6638           & 0.6843           & 0.6829  & 0.2623          
\\ 
\hline
KRED$_{BERT}$ single-task  & 0.6858  & 0.6931           & 0.6815           & 0.6963           & 0.6858         & 0.6714           & 0.6907  & 0.6885      & 0.2667      
\\ 
\hline
KRED$_{BERT}$              & \textbf{0.6885}         & \textbf{0.6937}  & \textbf{0.6916}  & \textbf{0.6984}  & \textbf{0.6869}           & \textbf{0.6768}  & \textbf{0.6910}           & \textbf{0.6914}  & \textbf{0.2684}  \\ 
 \specialrule{1.5pt}{1pt}{1pt}
\end{tabular}
\end{table*}

\noindent\textbf{Metrics}. For user2item task, we exploit \textsl{AUC} and \textsl{NDCG@10} as the evaluation metrics. For the item-to-item task, we use NDCG and Hit rate (Hit measures whether the positive item is present in the top-K list). For news popularity prediction and news category classification, we adopt \textsl{accuracy} and \textsl{macro F1-score} \footnote{\url{https://en.wikipedia.org/wiki/F1_score}} due to they are multi-class classification problems. For local news detection, because it is a binary classification problem, we adopt \textsl{AUC}, \textsl{accuracy}, and \textsl{F1-score} for evaluation. 

\noindent\textbf{Hyper-parameters}. Some hyper-parameter settings: the \textsl{learning rate} is 0.001, $L_2$ regularization coefficient $\lambda$ is set to 1e-5. The number of dimensions for hidden layers in the neural network is 128. The embedding dimension for entities is 90. The training process will terminate if the AUC on the validation set no longer increases for 10 epochs. The optimizer is \textsl{Adam}. In the KGAT layer, the maximum neighborhood of an entity is set to 20. The program is implemented with \textsl{PyTorch}. We release the source code at \url{https://github.com/danyang-liu/KRED} and we provide some instructions on running KRED on the MIND dataset \cite{wu-etal-2020-mind}.

\subsection{User-to-item Recommendation}
\label{sec:perrec}
Table \ref{tab:overall_auc} reports the detailed results of different models on the personalization recommendation task. Considering that the test set includes 7 days, we list all models' AUC performance on each day in order to verify whether the performance improvement is consistent. The last column `Overall' indicates the model performance on the whole 7 days. We make the following observations:
\begin{itemize}[leftmargin=*]
    \item Wide \& Deep outperforms FM, which indicates a deep model is better than a shallow model in modeling complex feature interactions. However, Wide \& Deep only uses a fully-connected multi-layer neural network to learn feature interactions and it heavily relies on feature engineering. For news articles, NLU models can produce much better document representations than hand-craft features. Thus, models (including NAML, DKN, STCKA, ERNIE, BERT and KRED) which contain an NLU component to extract document representation can easily achieve better performance than both Wide \& Deep and FM.
    \item Knowledge entities are indeed very important for news recommendations. This can be verified from that on both LDA+DSSM and BERT settings, DV$_*$+entity results are significantly better than DV$_*$ results.   KRED$_{LDA+DSSM}$ and KRED$_{BERT}$ further improve the DV$_{LDA+DSSM}$+entity and DV$_{BERT}$+entity settings by a large margin, which means that our model fuses the knowledge more effectively than an attention + concatenation approach.
     \item DKN, STCKA and ERNIE are three different types of knowledge-aware document understanding models which can serve as strong baselines. They are better than  DV$_{LDA+DSSM}$, but worse than KRED$_{LDA+DSSM}$ in most of the case, and worse than  KRED$_{BERT}$ across all metrics. This comparison further demonstrates the effectiveness of our proposed model in fusing knowledge information into document representation. 
     \item BERT representation achieves good performance, and it is even better than some baseline models which are knowledge-aware. This indicates that pretraining from large-scale language corpus (and followed up task-specific finetuning) greatly benefits the understanding of documents. Since our model can take an arbitrary kind of document vector as base input, we can enhance the BERT representation and the resulting model, i.e., KRED$_{BERT}$, achieves the best performance. Large-scale pretraining is very time-consuming and expensive; our flexible framework can directly consume the pretrained models and can be easily adapted to any new pretrained models in the future.  
    \item The joint-training KREDs (including both  KRED$_{LDA+DSSM}$ and KRED$_{BERT}$) outperform the corresponding single-task training KREDs (including KRED$_{LDA+DSSM}$ single-task and KRED$_{BERT}$ single-task), and the trend is consistent over the 7 days. During our experiments we find that, the impressive improvement on personalized user2item recommendation mainly comes from the jointly training of user2item recommendations and item2item recommendations. One possible reason is that, item2item relationship provide some intrinsic information and regularization between articles, which is not easy to be captured by the user2item training process, which essentially is a  multiple-items-to-item training mode.
\end{itemize}

\begin{figure*}
\setlength{\abovecaptionskip}{0cm}
\setlength{\belowcaptionskip}{-0.1cm}
\centering   
\subfigure  {
 \label{fig:item2item_a}     
\includegraphics[width=0.35\textwidth]{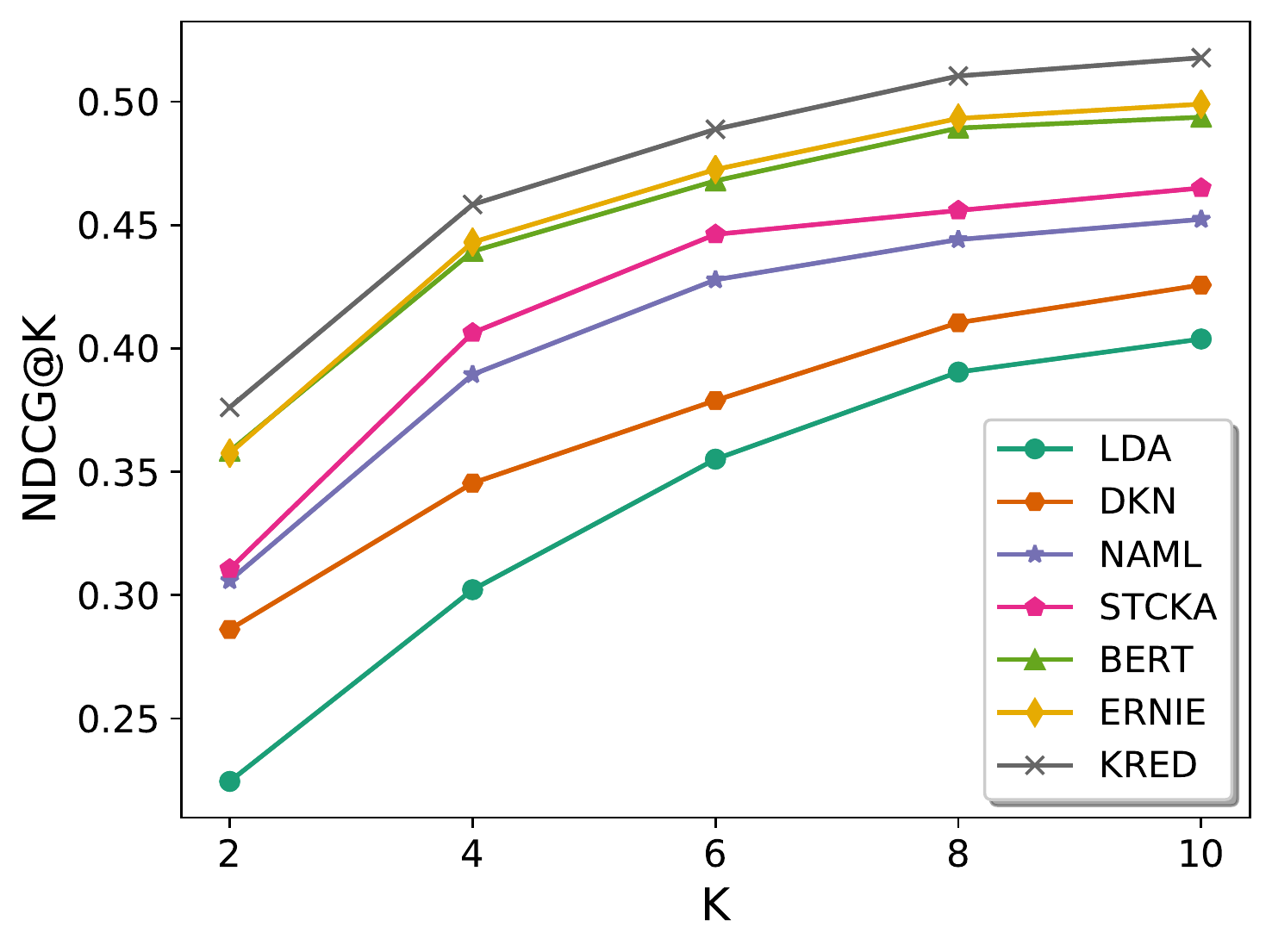}
}     
\subfigure  { 
\label{fig:item2item_b}     
\includegraphics[width=0.35\textwidth]{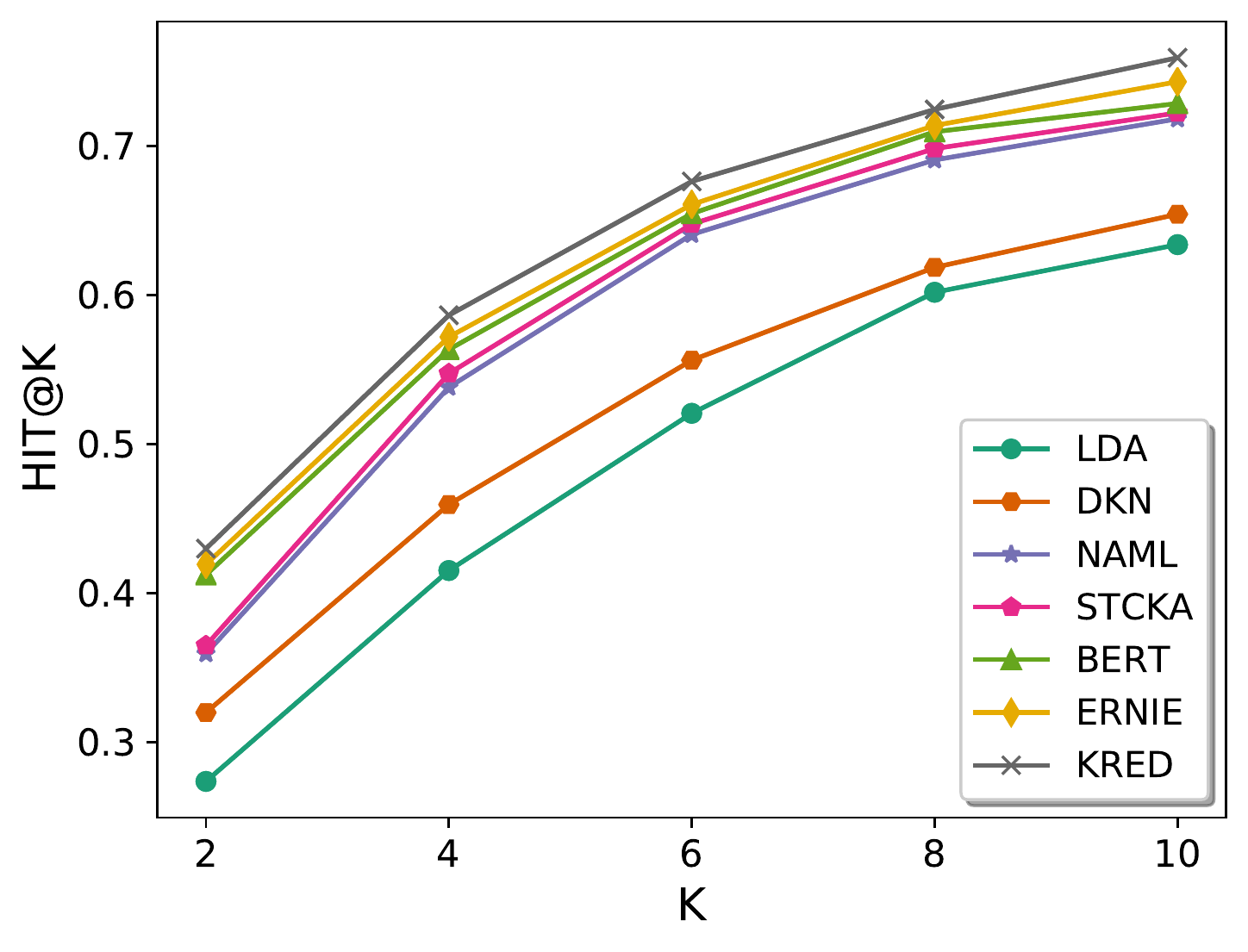}    
}   
\caption{Performance on the item-to-item task. Evaluation metrics are NDCG (left) and Hit ratio (right).}
\label{fig:item2item_results}
\end{figure*}

\subsection{Item-to-item Recommendation} 
\label{sec:item2item}
We construct an item-to-item recommendation dataset based on the original user-article clicking logs. The goal is to find a set of positive pairs $(a, b)$, such as that if a user has clicked on the article $a$, he/she will likely to click on the article $b$. We use co-click relationships to establish positive pairwise relations. 
If two news articles are clicked by more than 100 common users, we treat these two articles as a positive item pair. We randomly split them into train/valid/test set with the ratio of 80\%/10\%/10\%. For each item in the training and validation data, we randomly sample 5 negative news articles as the negative cases, and for the test dataset, we randomly sample 100 negative news articles in order to make the test evaluation more reliable.

In real-world industrial recommendation systems, approximate nearest neighborhood search (ANN) techniques are commonly used for fast item retrieval \cite{covington2016deep}. To use ANN techniques, items need to be encoded as a fixed-length vector. Thus for each model, we take its corresponding document representation vector, and use cosine similarity score to measure the relation between two documents. Figure \ref{fig:item2item_results} reports the results. All deep models outperform the LDA model with a large margin, which demonstrates the great power of deep learning techniques in text understanding. BERT and ERNIE take advantage of the pretraining process, so they are better than NAML. Our KRED further outperforms BERT and ERNIE, which further confirms that our proposed knowledge fusion technique is effective.

\begin{table*}
\setlength{\abovecaptionskip}{0.2cm} 
\caption{Performance comparison of different models on new article classification and prediction tasks. Numbers in \textbf{black bold} indicate that the result is significantly ($p$-value\textless 0.01) better than the best baseline method.}
\label{tab:documentcls}
\begin{tabular}{c|c|c|c|c|c|c|c}
  \specialrule{1.5pt}{1pt}{1pt}
\multirow{2}{*}{
} 
& \multicolumn{2}{c|}{Category Classification} & \multicolumn{2}{c|}{Popularity Prediction} & \multicolumn{3}{c}{Local News Detection}                                \\ \cline{2-8}  & ACC                   & F1-macro             & ACC                  & F1-macro            & \multicolumn{1}{l|}{AUC}             & ACC             & F1              \\ \hline
DKN                                  & 0.8148                & 0.6496               & 0.3775               & 0.3348              & \multicolumn{1}{l|}{0.8156}          & 0.9039          & 0.6258          \\ \hline
NAML                       & 0.8295                & 0.6933               & 0.3779             & 0.3412              & 0.8162                               & 0.9036          & 0.6263          \\ \hline 
STCKA                       & 0.8368                & 0.7209               & 0.3862               & 0.3504              & 0.8179                               & 0.9048          & 0.6264          \\ \hline 
ERNIE                           & 0.8602                & 0.7346               & 0.3887               & 0.3596              & 0.8255                               & 0.9145          & 0.6322          \\ \hline \hline
DV$_{LDA+DSSM}$                             & 0.8147                & 0.6473               & 0.3790               & 0.3352              & 0.8143                               & 0.9033          & 0.6242          \\ \hline
DV$_{LDA+DSSM}$+entity                             & 0.8232                & 0.6515               & 0.3840               & 0.3510              & 0.8179                               & 0.9080          & 0.6272          \\ \hline
KRED$_{LDA+DSSM}$                 & 0.8474                & 0.6924               & 0.3896               & 0.3641              & 0.8261                               & 0.9156          & 0.6453          \\ \hline \hline
DV$_{BERT}$                           & 0.8581                & 0.7246               & 0.3877               & 0.3592              & 0.8219                               & 0.9123          & 0.6317          \\ \hline
DV$_{BERT}$+entity                & 0.8592                & 0.7308               & 0.3898               & 0.3637              & 0.8245                               & 0.9135          & 0.6332          \\ \hline
KRED$_{BERT}$ single-task               & 0.8601                & 0.7336               & 0.3953               & 0.3725              & 0.8276                               & 0.9143          & 0.6357          \\ \hline
KRED$_{BERT}$               & \textbf{0.8703}       & \textbf{0.7651}      & \textbf{0.4031}      & \textbf{0.3900}     & \multicolumn{1}{l|}{\textbf{0.8418}} & \textbf{0.9176} & \textbf{0.6477} \\  \specialrule{1.5pt}{1pt}{1pt}
\end{tabular}
\end{table*}

\subsection{Document Classification and Popularity Prediction}
\label{sec:doccls}
The article category classification, article popularity prediction and local news detection tasks all belong to article-level classification problems. So we merge the performance comparison results into one table for concision. As reported in Table \ref{tab:documentcls}, the conclusions of the three tasks are consistent. Entity information is not only important for personalized recommendation and item-to-item recommendation, but also is very important for the three article-level classification tasks. Another interesting finding in Table \ref{tab:documentcls} is that the improvement of $KRED_{BERT}$ over $KRED_{BERT}$ single-task is impressive. The reason is that data labels for the three article classification tasks are very limited. However, the data from user-item interactions are much more abundant. Jointly considering multiple tasks in the training process can share and transfer knowledge across domains so different tasks can benefit from each other, and the benefit of the low-resource domain is especially significant.

 \begin{table}[t]
\setlength{\abovecaptionskip}{0.2cm} 
\caption{Ablation studies. Removing any of the layers in \textsl{KRED} leads to a performance drop.}
\label{tab:removelayer}
\begin{tabular}{l|c c|c c} \specialrule{1.5pt}{1pt}{1pt}
    Model & AUC       &     & NDCG@10          &                  \\ \hline
       KRED     & \textbf{0.6894}  &    &   \textbf{0.2673}    &                       \\  
        w/o KGAT          & 0.6853  &  $\downarrow$ 0.59\%  &  0.2642   & $\downarrow$  1.16\%     \\  
        w/o Context      & 0.6859  & $\downarrow$ 0.51\% & 0.2644    &      $\downarrow$ 1.08\%      \\         
        w/o Distillation   & 0.6845  & $\downarrow$ 0.71\%   &      0.2631   &   $\downarrow$ 1.57\%    \\  \specialrule{1.5pt}{1pt}{1pt} 
\end{tabular} 
\end{table}

\subsection{Ablation Study}
Next we investigate how each layer contributes to KRED. Since there are three key layers in KRED, we remove one layer each time and test if the performance is influenced. Table \ref{tab:removelayer} shows the results. Because the conclusions on different tasks are similar, we only report the results of KRED$_{BERT}$ on the personalized recommendation task to save space. We observe that removing either one layer will cause a significant performance drop, which indicates that all layers are necessary.

   \begin{table}[h]
       \caption{Average time cost (in seconds) of training/inferring one epoch for 100k documents} 
     \label{tab:timecomplexity} 
     \centering 
     \begin{tabular}{c c c}     \specialrule{1.5pt}{1pt}{1pt}
             method & training time (s) &  inference time (s) \\ \hline    
        KRED  &   10.52 & 5.17\\ \hline  
        KCNN     &   88.48 & 19.46\\ \hline  
        STCKA    &   93.27 & 21.58\\ \specialrule{1.5pt}{1pt}{1pt}
     \end{tabular} 
 \end{table}
\subsection{Efficiency Comparison}
\setlength{\abovecaptionskip}{0.2cm}  
 To evaluate the efficiency, we compare the computational cost of our KRED model with STCKA and the KCNN module in DKN. ERNIE belongs to a pretrained model like BERT so it is not suitable for comparison. Experiments are conducted on a Linux machine with GPU Tesla P100, CPU Xeon E5-2690 2.60GHz 6 processors. We prepare 100,000 documents and count the average time cost for training/testing one epoch of each model. The batch size is set to 64. The maximum word length is set to 20 for the title and 1000 for the body. The results are shown in Table \ref{tab:timecomplexity}. Unlike KCNN and STCKA which model the whole text content, KRED takes the aggregated information of entities as additional input, so it is much faster in both training and inference steps.

\subsection{Visualization of Document Embeddings}
\begin{figure*}
\centering   
\includegraphics[trim=0 10 0 10,clip,width=1.0\textwidth]{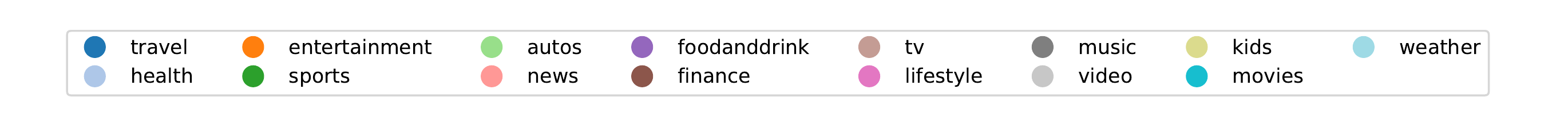}
\subfigure[KRED. C-H score = 426] {
 \label{fig:a}     
\includegraphics[trim=10 10 0 35,clip,width=0.32\textwidth]{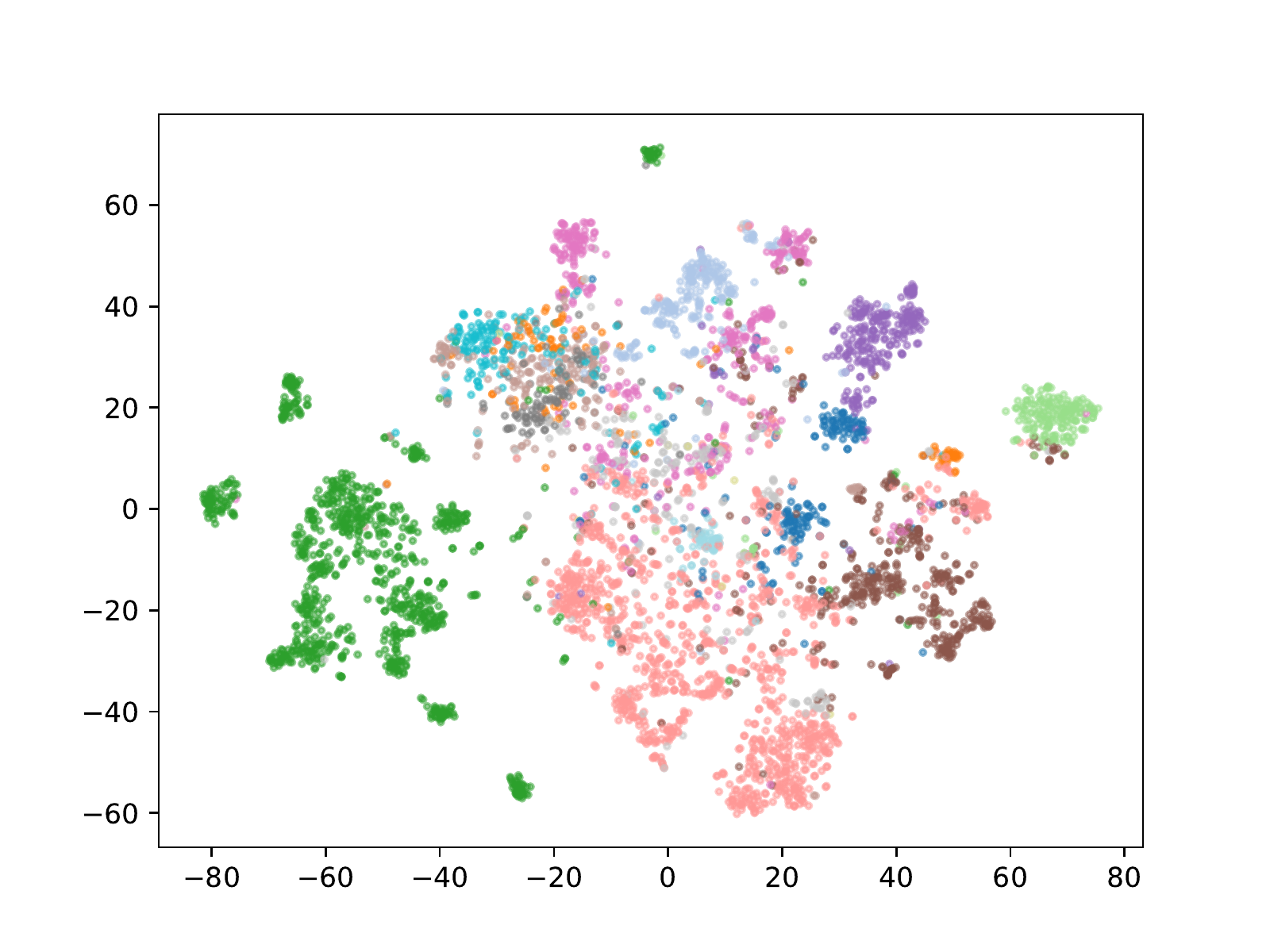}
}\hspace{-7mm}
\subfigure[BERT. C-H score = 375] { 
\label{fig:b}     
\includegraphics[trim=10 10 0 35,clip,width=0.32\textwidth]{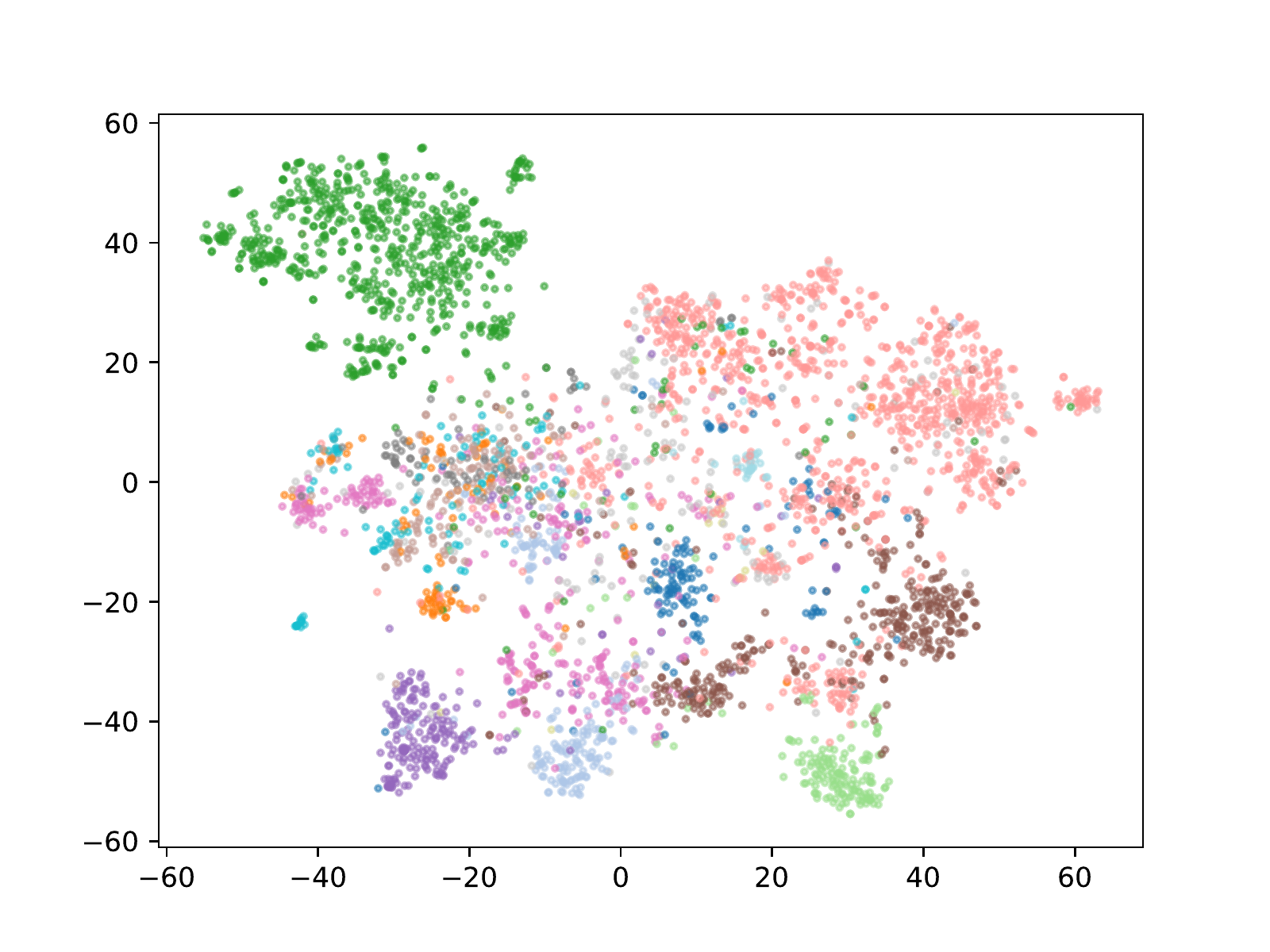}    
}\hspace{-7mm}  
\subfigure[LDA. C-H score = 68] { 
\label{fig:c}     
\includegraphics[trim=10 10 0 35,clip,width=0.32\textwidth]{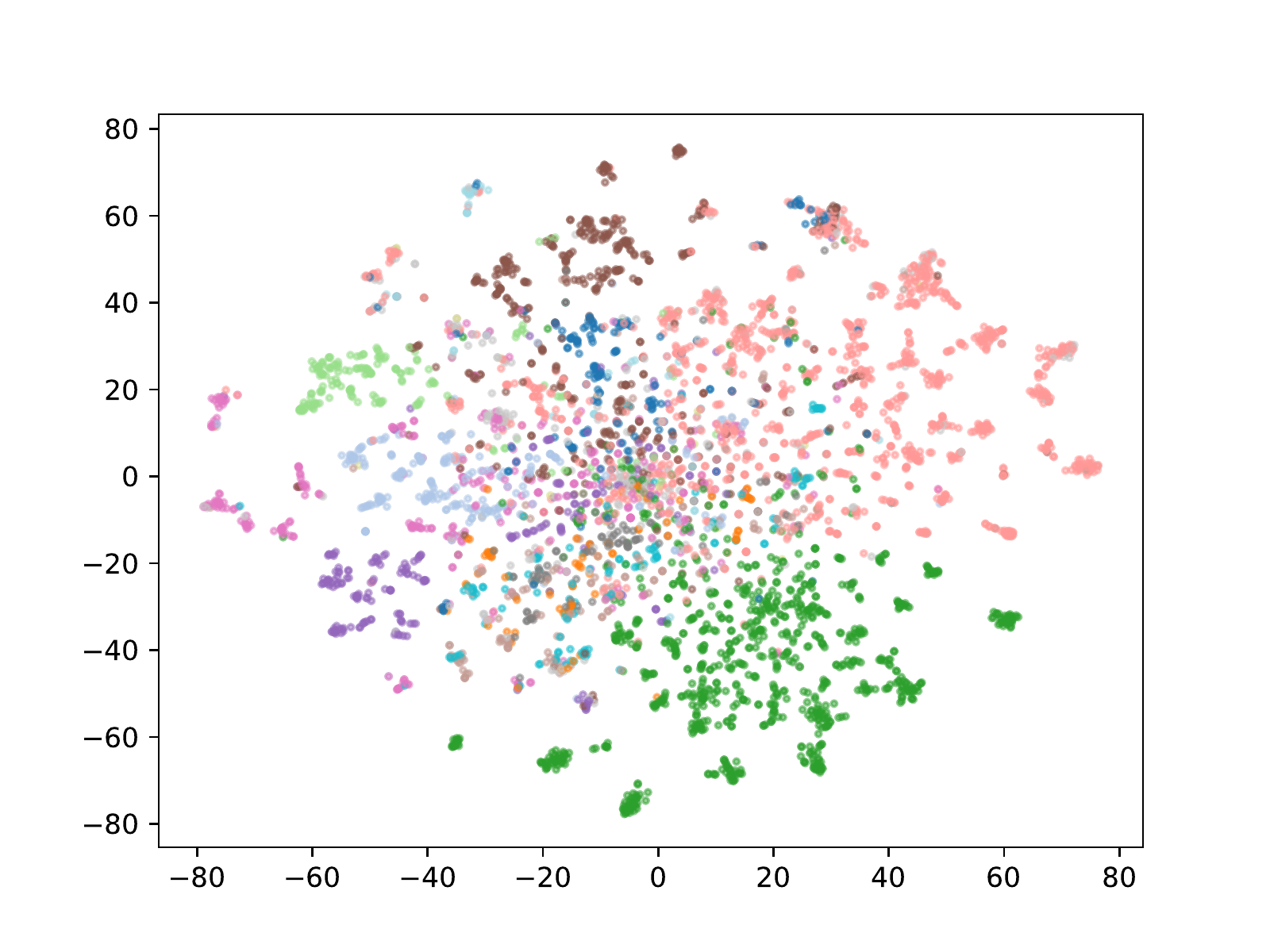}  
}
\caption{t-SNE visualizations of document embeddings of KRED, BERT and LDA. Color indicates the category of a document.}
\label{fig:knowledge_fusion_architectures}
\end{figure*}

To better understand the quality of document representations, we conduct a visualization study on document embeddings. Specifically, we first project the document representations into a 2-D space with the t-SNE algorithm \cite{maaten2008visualizing}, then we plot the distribution of documents in Figure \ref{fig:knowledge_fusion_architectures}, where each point represents a document, and documents are denoted with different colors based on their categories. We can observe that documents with the same category tend to cluster together, and the embeddings produced by the KRED model present a better distribution pattern than that of the BERT and LDA model. We further compute the  Calinski-Harabaz score \cite{calinski1974dendrite} as a quantitative indicator to compare the clustering patterns of different models. A higher Calinski-Harabasz score relates to a model with better defined clusters, and the scores for KRED, BERT and LDA models are 426, 375 and 68, respectively.

\subsection{Case Study}
\begin{figure*}
 \vspace{-0.05cm}
 \setlength{\abovecaptionskip}{-0.005cm}
    \centering
    \includegraphics[width=0.95\textwidth,trim={0 0cm 0 0},clip]{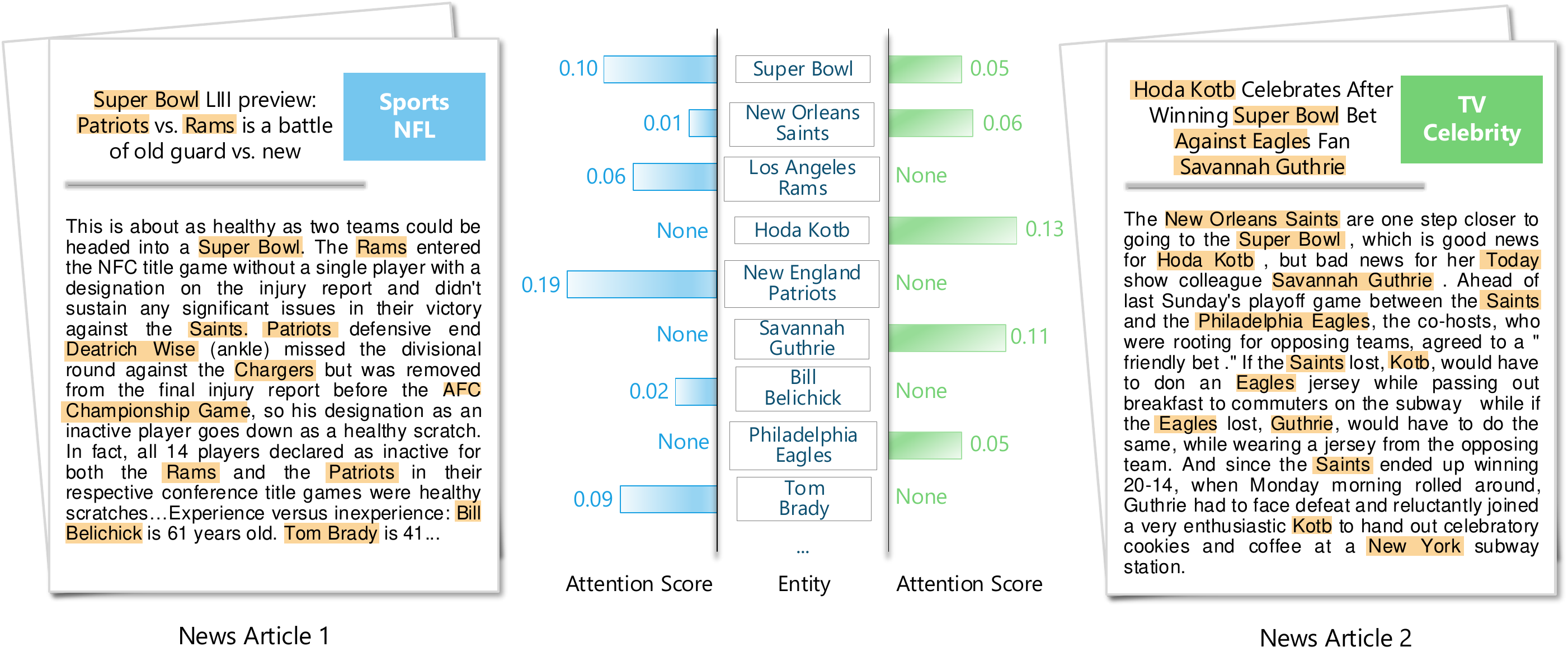}%
    \caption{A case study on what attention scores the KRED model will assign to different entities, and how a same entity will take different importance scores in different documents. None means the entity doesn't appear in the article. }
    \label{fig:case study}
\end{figure*}

We provide a case study on what attention scores the KRED model will assign to different entities, and how a same entity will take different importance scores in different documents. We select two articles which have several entities in common. As depicted in Figure \ref{fig:case study}, a news article may mention a dozen of entities, while only a few of them are key entities. In Article 1, whose category is \textsl{Sports NFL}, the key entities are \textsl{Super Bowl}, \textsl{Los Angeles Rams}, \textsl{New England Patriots}, because the topic of Article 1 is about a preview of Super Bowl LIII between the two teams. Although some other teams, i.g. \textsl{New Orleans Saints}, are mentioned in the article, their attention scores are relatively much lower.  Article 2 is a piece of news reporting a bet between two famous hosts, \textsl{Hoda Kotb} and \textsl{Savannah Guthrie}. The bet is about two sports teams, \textsl{New Orleans Saints} and \textsl{Philadelphia Eagles}. So the entity \textsl{New Orleans Saints} is assigned with a higher attention score in Article 2 than in Article 1.

\section{Related Work}
\subsection{News Recommender Systems}
\cite{Okura:2017:ENR:3097983.3098108} argues that ID-based methods are not suitable for news recommendations because candidate news articles expire quickly. The authors exploit a denoising autoencoder to generate representation vectors for news articles. \cite{Lian:2018:TBR:3304222.3304298} explores how to effectively leverage content information from multiple diverse channels for better news recommendations. \cite{wu2019npa} proposes an attention model which can exploit the embedding of user IDs to generate personalized document representations. \cite{DBLP:conf/ijcai/WuWAHHX19} takes news titles, bodies and categories as different views of news, so it introduces a multi-view learning model to learn news representations. \cite{wu2019npa} further proposes to learn personalized news representation by integrating user embeddings in the news representation learning process. However, the aforementioned works either use hand-craft features or extract representations from text strings; none of them study the usage of knowledge graphs.  DKN \cite{Wang:2018:DDK:3178876.3186175}, which takes a knowledge graph as side information, is the most related work to this paper, we have stated the relationships between DKN and our work in the introduction section, and we report the performance comparison in the experiment section. \cite{ijcai2019-424} summarizes the user history into multiple complementary vectors in the offline stage, and it reports the advantage of this method on a news recommendation scenario. \cite{DBLP:conf/cikm/LiuBLZSWX19} constructs a domain-specific knowledge graph, called news graph, for better knowledge-aware news recommendations.
\subsection{Knowledge-Aware Recommender Systems}
Incorporating a knowledge graph as side information has proven promising in improving both accuracy and explainability for recommender systems. \cite{Cao:2019:UKG:3308558.3313705} jointly learns the model of recommendation and knowledge graph completion, which achieves better recommendation accuracy and a better understanding of users' preferences. \cite{wang2019multi} proposes a multi-task feature learning approach for knowledge graphs enhanced recommendation.  \cite{Wang:2019:KKG:3292500.3330989} argues that traditional methods such as FM assume each user-item interaction is an independent instance, while overlooks the relations among instances or items. The authors thus propose using knowledge graphs to link items with their attributes. \cite{wang2018ripplenet} introduces an end-to-end framework, named \textsl{RippleNet}, which simulates the propagation of user preferences over the set of knowledge entities and alleviates the sparsity and cold start problem in recommender systems. \cite{Wang:2019:KGN:3292500.3330836,wang2019knowledge} study the utilization of graph convolutional networks as the back-end model for incorporating knowledge graph and recommender systems. However, these knowledge-aware recommendation models cannot be directly applied to news recommendations, because a news article cannot be mapped to a single node in the knowledge. Instead, it is related to a list of entities, effectively fusing knowledge entities into document content is the key to successful news recommendations.

\section{Conclusions}
We propose the KRED model, which enhances the representation of news articles with entity information from a knowledge graph in a fast, elastic, accurate manner. There are three core components in KRED, namely the entity representation layer, the context embedding layer and the information distillation layer. Different from most existing works that only focus on the personalized recommendation task, we study five key applications for news recommendation service, including personalized recommendation, item-to-item recommendation, local news detection, news category classification and news popularity prediction, and propose to jointly optimize different tasks in a multi-task learning framework. Extensive experiments demonstrate that our proposed model consistently and significantly outperforms baseline models. For future work, we will study a \textsl{KREU} model, which means a Knowledge-aware Representation Enhancement model for Users.

\bibliographystyle{ACM-Reference-Format}
\bibliography{kred.bib}


\end{document}